\documentclass[preprint]{aastex}

\slugcomment{Submitted to the Astronomical Journal}

\begin{document}

\title{Discovery of a High-Redshift (z=0.96) Cluster of Galaxies
Using a {\it FIRST} Wide-Angle Tailed Radio Source\altaffilmark{1}}

\author{Elizabeth L. Blanton\altaffilmark{2,3,4},
Michael D. Gregg\altaffilmark{5,6},
David J. Helfand\altaffilmark{4,7},
Robert H. Becker\altaffilmark{5,6},
and Richard L. White\altaffilmark{8}}

\altaffiltext{1}{Based in part on observations obtained at the W. M. Keck
Observatory.}

\altaffiltext{2}{{\it Chandra} Fellow}

\altaffiltext{3}{Department of Astronomy, University of Virginia,
P. O. Box 3818, Charlottesville, VA 22903;
eblanton@virginia.edu}

\altaffiltext{4}{Visiting Astronomer, Kitt Peak National Observatory,
National Optical Astronomy Observatories, which is operated by the
Association of Universities for Research in Astronomy, Inc., under
cooperative agreement with the National Science Foundation}

\altaffiltext{5}{Department of Physics, University of California at Davis,
1 Shields Avenue, Davis, CA  95616;
gregg@igpp.ucllnl.org,bob@igpp.ucllnl.org}

\altaffiltext{6}{Institute of Geophysics and Planetary Physics, Lawrence
Livermore National Laboratory, L-413, P.O. Box 808, Livermore, CA  94550}

\altaffiltext{7}{Columbia Astrophysics Laboratory, 550 West 120th St., 
New York, NY  10027; djh@astro.columbia.edu}

\altaffiltext{8}{Space Telescope Science Institute, 3700 San Martin Dr.,
Baltimore, MD 21218; rlw@stsci.edu}

\begin{abstract}
Using a combination of near-infrared and optical photometry, along with
multi-object spectroscopy, we have confirmed the existence of a 
high-redshift cluster of galxies at $z = 0.96$.  The cluster was found using
a wide-angle tailed radio source selected from the VLA {\it FIRST} survey as
a cluster signpost.  These types of radio sources are often found in clusters,
and are thought to attain their C-shaped morphologies from the relative 
motion between the radio source host galaxy and the intracluster medium.
We present optical/near-infrared color-magnitude diagrams which show a
concentration of cluster galaxies in color space.  We also include 
spectroscopic results obtained from the Keck II LRIS.  Ten galaxies are 
confirmed at the cluster redshift, with a line-of-sight velocity dispersion
of $\sigma_{\|}=530^{+190}_{-90}$ km s$^{-1}$, typical of an Abell richness
class 0 cluster. Using data from the {\it ROSAT} public archive, we limit the
X-ray luminosity for the cluster to 
$L_{X,bol} \la 3 \times 10^{44}$ erg s$^{-1}$, consistent with the value
expected from the $L_{X}-\sigma$ relation.

\end{abstract}

\section{Introduction}

Clusters of galaxies trace the largest gravitationally-bound mass 
concentrations in the
Universe; their epoch of formation, and thus their abundance as a function of
redshift, is a strong constraint on $\Omega_M$.  
High-redshift clusters are very 
difficult to find -- the faint optical magnitudes of the associated galaxies
and the X-ray emission from the intracluster medium (ICM) will often 
be missed by current surveys at redshifts of $z\approx1$.  
Any new clusters revealed at these distances add
significantly to the few that are known, and are useful for studies of 
cosmology and galaxy evolution.

Using a unique method based on radio galaxy morphology, we have confirmed 
the existence of dozens of 
clusters with redshifts up to $z=0.85$ (Blanton et al.\ 2000, 2001).  
Bent double-lobed radio sources (particularly FR I-type sources 
[Fanaroff \& Riley 1974], including wide-angle tails [WATs] and 
narrow-angle tails [NATs]) are thought to achieve their distinct 
``swept-back'', C-shaped 
morphologies as a result of interaction with dense gas in which they are 
embedded.  The canonical model describes the radio source's host galaxy as 
having a significant peculiar velocity relative to the cluster with which it 
is associated -- the ram pressure of the dense ICM then pushes back the lobes
of the radio source giving it a ``bent'' appearance (Owen \& Rudnick 1976).  
Often, the host galaxies of WAT radio sources are cD ellipticals, thought
to reside at the bottom of cluster potential wells, and are thus not expected
to have significant peculiar velocities.  An alternate explanation 
(Burns et al.\ 1994), posits
bulk motion of the ICM as a result of a recent or ongoing cluster-cluster
(or sub-cluster) merger.  The ram pressure of the merging gas then bends the
lobes.  Either way, bent-double radio sources are often pointers to 
clusters, and they may be flags in particular for clusters that have 
recently formed or merged.

We have selected 384 bent-double (mostly WAT) radio sources from the first 
3000 deg$^2$ of
the VLA {\it FIRST} (Faint Images of the Radio Sky at Twenty-cm) survey 
(Becker, White, \& Helfand 1995).  Imaging
and multi-slit mask spectroscopic follow-up on a moderate-to-high redshift 
subset of ten of these 
revealed eight clusters with redshifts as high as $z=0.85$ 
(Blanton et al.\ 2000).
A complete (area- and magnitude-limited) low-redshift subset including 
40 of the 384 objects was observed
optically, and $\sim50\%$ of them were found to be in richness
class Abell 0 or greater clusters (Blanton et al.\ 2001).  The majority of 
the low-redshift bent doubles that are found in clusters, based on our 
optical richness measurements, are detected in the ROSAT All Sky Survey 
(five out of six detected for $z<0.2$). Those that are not found in clusters, 
based on our 
richness measurements, are not detected in the RASS (only one out of nine 
detected for $z<0.2$).  In other words, the
combination of radio and optical data gives us a very clear indication for
the presence, or lack of, a cluster, as confirmed with X-ray observations.

In order to identify high-$z$ candidates, we obtained $R$-band images of fields
surrounding bent doubles that were blank to the limit of the Digitized 
Sky Survey (DSS).  Some of these had no optical identifications even down to the 
limit of our images ($m_{R}\sim22-23$).  We then performed near-infrared (NIR)
observations at the positions of these sources in hopes of uncovering
very red, high-$z$ cluster galaxies.  

NIR observations have been successful in
the past in identifying high-$z$ clusters. Stanford et al.\ (1997, 2001) 
found one of the most distant clusters known at $z=1.27$ using NIR and optical
imaging, while
Rosati et al.\ (1999) used a combination of X-ray, NIR, and optical 
observations to confirm
the existence of a cluster at $z=1.26$ that may be part of a superstructure
including the Stanford et al.\ (1997) cluster.
Using the NIR alone, or in conjunction with radio and/or X-ray
data, gives the potential for locating clusters that are at such high-$z$ 
they would be missed by all but the deepest optical observations.
Recently, Gladders \& Yee (2000) have performed a high-redshift cluster 
search by using optical and NIR photometry and identifying clusters by their
``red sequence,'' or clustering on a color-magnitude diagram.  Chapman et
al.\ (2000) have used a combination of radio observations with optical and
NIR photometry to identify likely clusters at redshifts greater than 0.8.

Here we present optical and NIR observations taken at the position of the 
bent-double radio source, BD1137+3000.  A combination of imaging and 
spectroscopy confirms that this source is located in a distant cluster of
galaxies with a redshift $z=0.96$.

We assume $H_{\circ} = 70$ km s$^{-1}$ Mpc$^{-1}$, $\Omega_M = 0.3$,
and $\Omega_{\Lambda} = 0.7$ throughout.

\section{BD1137+3000:  Radio Data}

\begin{figure}
\plotone{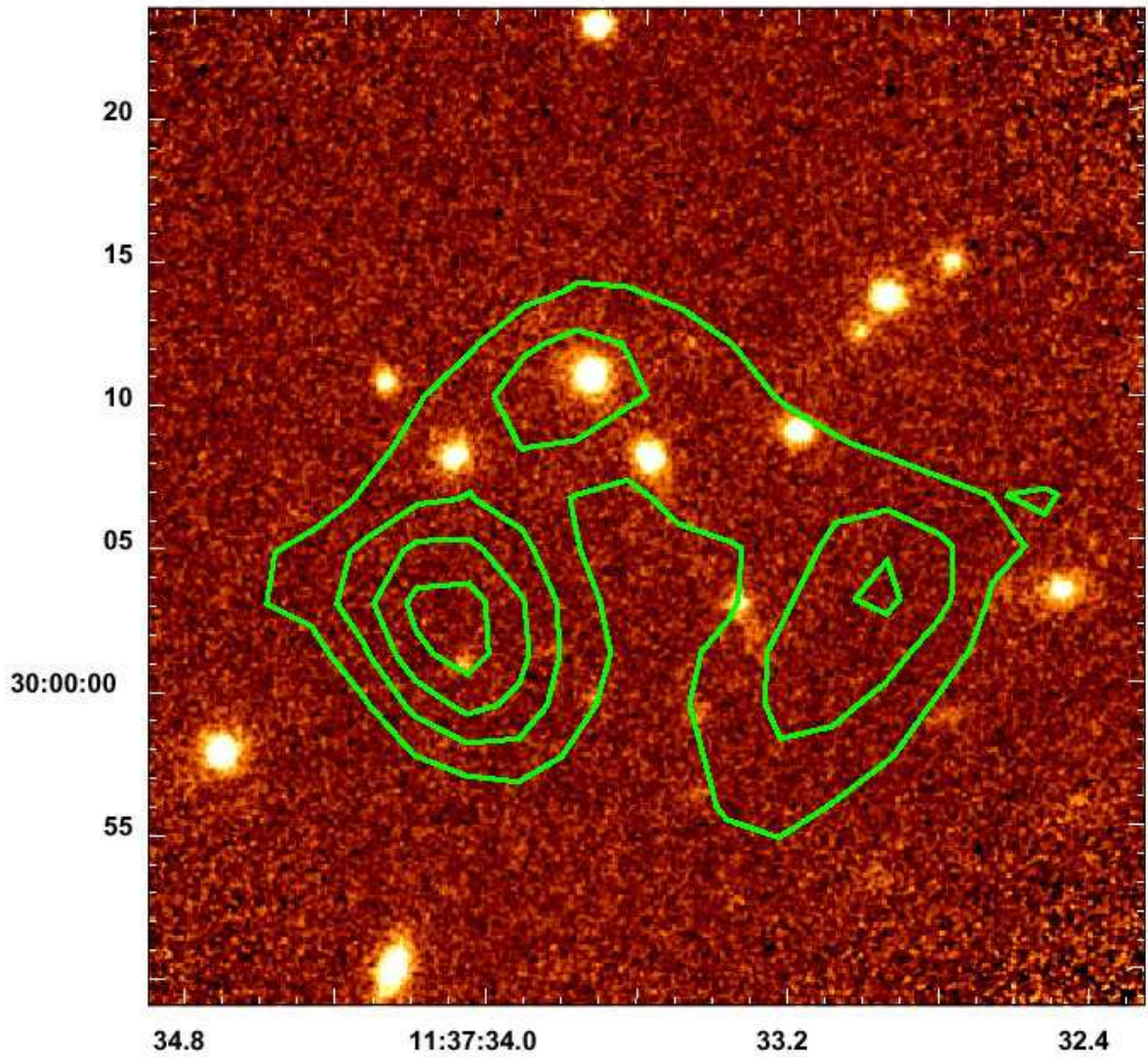}
\caption{Overlay of the {\it FIRST} 20cm radio contours onto a 
$K^{\prime}$-band image taken with the Keck NIRC.  The contours are
logarithmically spaced in the range 0.45 to 4.0 mJy.  With our assumed
cosmology, the figure is $280\times280$ kpc.\label{fig:ovly}}
\end{figure}

The radio source is detected in the {\it FIRST} survey as a three component
object, including a core and two lobes (FIRST J113733.6+300009,
FIRST J113734.0+300002, and FIRST J113732.8+300003).  The position of the 
core is
$\alpha$ = 11h 37m 33.62s, $\delta = +30^{\circ} 00\arcmin 9\farcs5$.
The source is clearly bent with an opening angle between the two lobes of 
$\sim 80^{\circ}$.  The size of the source, measured from the outer edge of 
one lobe
to the core and out to the outer edge of the other lobe is $30\arcsec$
(240 kpc).  The total flux density, as
measured by {\it FIRST} is 9.1 mJy.  The southeast lobe is the brightest of
the three components, with a flux density of 4.1 mJy.  The core has a 
comparable
flux density (3.8 mJy), while the southwest lobe is 1.2 mJy.  It is
difficult to classify the source as an FR I or II based on its morphology.
The southeast lobe is bright at the end of the lobe (typical of an FR II
source).  The southwest lobe brightens approximately halfway between the
core and the edge of the lobe, bends at this bright area, and then fades
toward the edge of the lobe (typical of an FR I source).  We expect that
there is additional extended emission related to the source that is not 
seen because (1) there is surface brightness dimming $\propto (1+z)^4$ and (2)
it is resolved out by the high resolution of the {\it FIRST} images.
The flux density measured from the low-resolution
NRAO VLA Sky Survey ({\it NVSS}; Condon et al.\ 1997), which resolves out 
less extended emission, 9.8 mJy, is slightly higher than the {\it FIRST}
value. Using the {\it NVSS} flux density value, and assuming 
$P \propto \nu^{-\alpha}$ with $\alpha = 0.8$, the radio power of the 
source at 1440 MHz is $P_{1440} = 4.0 \times 10^{25}$ W Hz$^{-1}$.  
This power is slightly below the FR I/II break for radio galaxies 
(Fanaroff \& Riley 1974), and is typical of the class of bent double radio 
sources known as wide-angle tails (WATs; O'Donoghue et al.\ 1993) which are
very often associated with cD galaxies in the centers of clusters. 
An overlay of the {\it FIRST} radio contours onto a Keck near-infrared
camera (NIRC) image in the K$^{\prime}$ filter of the cluster
center is shown in Figure~\ref{fig:ovly}.  The figure is 280 kpc on a side
for our assumed cosmology.  The radio contours are 
logarithmically spaced in the range 0.45 to 4.0 mJy.  The radio core is
coincident with an elliptical galaxy, as seen in the near-infrared image.
This NIR image, which covers only the inner $35\arcsec$ square region of the 
cluster, is used for display purposes only in this paper.  We include analysis
of NIR data from the KPNO 2.1m telescope covering a much wider field of view
in the following section.

\section{Optical/NIR Imaging Data and Reductions}

Optical data taken in the R-band at the KPNO 2.1m under partly cloudy 
conditions on UT 6 April 1997 revealed no identification for the radio 
galaxy to
a limit of $R \approx 22$.  This motivated us to observe the object in
the NIR, and later to reobserve it to deeper limits in the optical  
under photometric conditions.

\subsection{ NIR}
BD1137+30 was observed with IRIM at the KPNO 2.1m on UT 25 and 26 May 1997;
the image scale was $1\farcs09$/pixel.
Observations were taken in a 16-point grid pattern.  The total exposure
in the $K$-band was 64 minutes (64 frames of 4$\times$15 s coadds), and that
in the $J$-band was 32 minutes (16 frames of 4$\times$30 s coadds).  The 
seeing was approximately $1\farcs6$ for the $J$-band image and
$1\farcs7$ for the $K$-band image.  A non-linearity correction
was applied to the data to account for the changing response of the 
instrument when
high fluxes are encountered.  The correction was kindly provided by 
F. R. Chromey, derived from a non-linearity sequence taken the night 
immediately
following our run, and amounted to a 3.3\% correction at 30,000 counts.
The data were corrected for the
effects of dark current and  flat fields using IRAF.  The flat fields were
constructed for each band by median combining the dithered target images
and rejecting the objects contained therein.  These types of flats were
superior to dome flats (which we also constructed).  The images were then 
processed with the IRAF contributed package 
DIMSUM\footnote{Deep Infrared Mosaicking Software, um...  Created by
P. Eisenhardt, M. Dickinson, S. A. Stanford, \& J. Ward}.  We used DIMSUM to
sky-subtract the data by using the sky derived from eight images taken 
immediately
before and after each frame.  The frames were then aligned and combined to
produce a 'first pass' combined image.  During this first pass, object
masks were constructed for the frames to help eliminate object residuals in
the final combined image.  A 'mask pass' run incorporated these masks and
produced the final combined image.  We modified DIMSUM to attain better
alignment of the images before combining them.  The final mosaicked image
reduced the image scale by a factor of four (to $0\farcs27$/pixel --
see Stanford, Eisenhardt, \& Dickinson (1995)).  During the reduction 
process, a few bad frames were identified and rejected,
leaving us with a total of 30 min in the $J$-band and 56 min in the $K$-band.
The final 'mask pass' combined images were normalized so that the counts
are counts per second.  UKIRT faint standards were used to derive the
photometry.

\subsection{Optical}
A total exposure in the $R$-band of 7200 s ($6 \times$ 1200s) at the MDM 1.3m 
telescope with
the Templeton CCD obtained on UT 21 May 1998 was kindly provided by J. Kemp.
This setup produced an image scale of $0\farcs5$/pixel.  The night
was photometric and the seeing was approximately $1\farcs2 -
1\farcs3$.  The data were processed following standard IRAF 
procedures to remove the effects of the bias and flat field.  The images
were aligned and median-combined with cosmic rays rejected.  The Landolt
standard field PG1323 was used to calibrate the data.

\section{Photometry / Image Analysis}
FOCAS was used to measure magnitudes for the standard stars and objects
in the target fields for the NIR and optical data.  FOCAS 'total' magnitudes 
were measured for the standards.
The error for the optical photometry, estimated from the scatter in the
offset between the measured and Landolt-catalogued magnitudes for five stars
in PG1323, is 0.04 mag.  The error in the NIR magnitudes, estimated
from the difference in the zero point derived for the two nights, is
0.1 mag.

Aperture magnitudes were measured for objects in the target field.  To
assure that we were measuring the same physical aperture in each band, as
well as to make the analysis more straightforward,
adjustments were made to the final $R$, $J$, and $K$ images.  They were all
scaled to have the same pixel scale ($0\farcs5$/pix) and the
same orientation (N up, E to the left).  The $R$ and $J$ images were smoothed
with a Gaussian to match the seeing in the $K$-band image (the worst of the 
three, with 1$\farcs7$).  We then used FOCAS to measure magnitudes
within an aperture with a diameter equal to 2$\times$FWHM
(3$\farcs4$ = 6.8 pixels).  This is close to a total 
magnitude and will measure the same part of the galaxy in each of the bands.
At $z=0.96$ (see \S\ref{sec:spect}), this aperture corresponds to 27 kpc
for our assumed cosmology.  The detection
limit was 2.5$\sigma$ above sky and the minimum area was 12 
pixels$^2$ = 2.9 arcsec$^2$ (the square of
the FWHM).  The sky was calculated by taking the average pixel value in
a square annulus (default width 3 pixels) some buffer distance (default 2 
pixels) away from the edge of the objects' last isophotes.  A small difference
in the sky value had a very significant effect on faint objects.  To assure 
that
we weren't including galaxy light in the sky subtraction, since many of
the galaxies are very close together, we increased the buffer distance
until the change in distance no longer affected the measured aperture
magnitude (i.e., the sky value was no longer changing).  We used a buffer
of 15 pixels in $R$ and 20 pixels in $J$ and $K$. 

The aperture magnitudes, measured as described above, for the radio source 
counterpart are 22.90$\pm0.04$, 19.46$\pm0.1$, and 17.78$\pm0.1$, in $R$, 
$J$, and $K$, respectively.
This gives colors of ($R-K$) = 5.12$\pm0.11$, ($J-K$) = 1.68$\pm0.14$, and 
($R-J$) = 3.44$\pm0.11$.
The absolute magnitude of the host galaxy is $M_{V} = -22.29$, assuming no
evolution, a redshift $z=0.96$ (see \S\ref{sec:spect}), and a $K$-correction
from Coleman, Wu, \& Weedman (1980).  This is
typical of a bright elliptical galaxy, but fainter than a cD.

\section{Color-Magnitude Diagrams}

\begin{figure}
\plotone{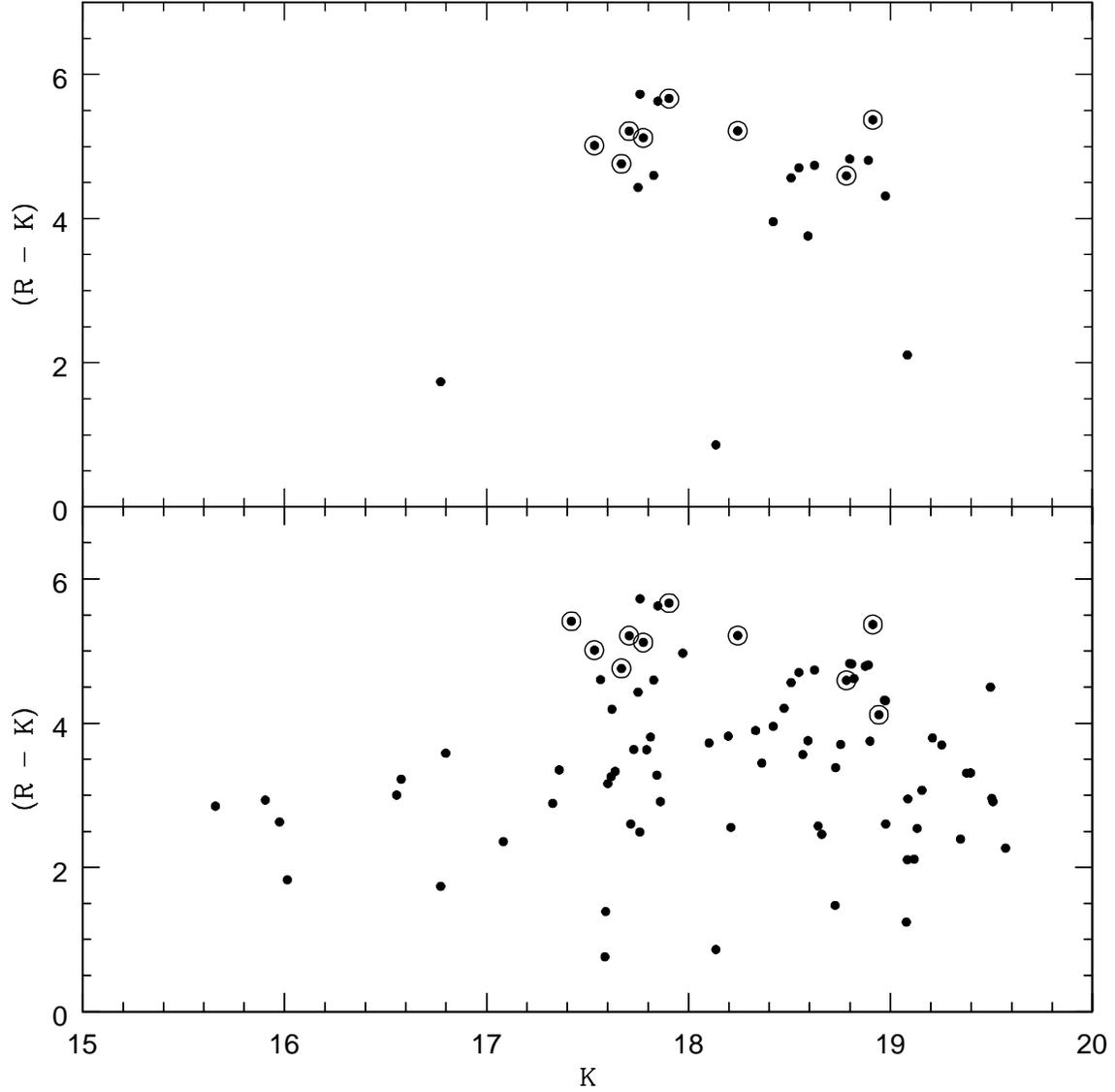}
\caption{$K$ vs. $(R-K)$ color-magnitude diagram.  The top panel includes 
the objects within a $1\arcmin$ radius from the radio source, while the bottom
panel includes the objects in the entire 15.6 arcmin$^2$ field.  There is
a concentration of objects with $4 < R-K < 6$ that are likely cluster
members. Spectroscopically confirmed cluster members are circled.\label{fig:krk}}
\end{figure}

\begin{figure}
\plotone{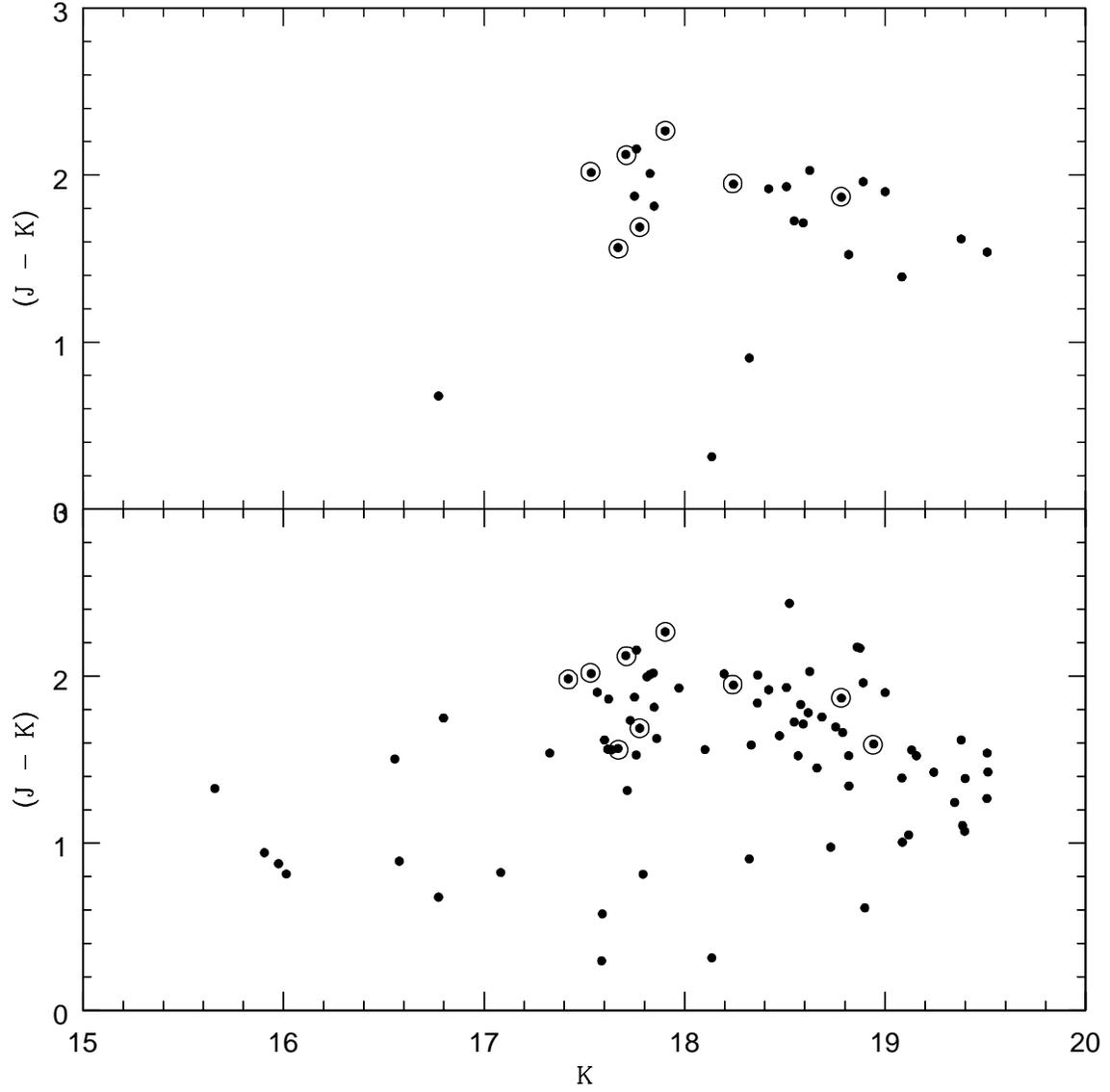}
\caption{$K$ vs. $(J-K)$ color-magnitude diagram.  As in 
Figure~\ref{fig:krk}, the top panel includes only the
objects within a $1\arcmin$ radius from the radio source, and the bottom
panel includes the objects in the entire 15.6 arcmin$^2$ field.  Probable
cluster members have approximately $1.5 < J-K < 2.5$.  Spectroscopically
confirmed members are circled.
\label{fig:kjk}}
\end{figure}

Color magnitude diagrams (CMDs) were constructed for objects in the entire 
15.6 arcmin$^2$ field as well as for only the objects found within a 
$1\arcmin$ radius aperture (480 kpc at $z = 0.96$, see \S\ref{sec:spect}) 
from the radio source.
No separation was made between stars and galaxies (all were included).  At
magnitudes close to that of the radio source, star/galaxy separation is
very difficult, and not very important since number counts at these faint 
magnitudes are dominated by galaxies.  The CMDs are presented in 
Figures~\ref{fig:krk}
and \ref{fig:kjk}.  The top panels include the objects within $1\arcmin$ from
the radio source, while the bottom panels include the objects from the
entire field. The objects that are
confirmed to be at the cluster redshift (see \S\ref{sec:spect}) are circled.
Three obvious, foreground stars are seen in the 
diagrams -- the three blue objects apparent in the CMDs constructed from the 
objects within one arcmin of the radio source.
There is a much larger range in the colors of objects when looking at the 
CMDs for the whole field.
The objects in the $1\arcmin$ radius aperture have less scatter (they are
much more concentrated on the diagram) than
those in the whole field, and are redder, consistent with membership in
a distant cluster.  Most of the objects within $1\arcmin$ from the radio
source have colors in the range $4 < (R-K) < 6$ and 
$1.5 < (J-K) < 2.5$.
Comparing the CMDs to those presented in Stanford, Eisenhardt, \&
Dickinson (1998), which includes diagrams for 19 clusters from 
$z = 0.3 - 0.9$,
and Stanford et al. (1997), which displays a CMD for a
$z = 1.27$ cluster, suggests that this cluster has a redshift between 
$z = 0.9$ and 1.27, and closer to 0.9.

\begin{figure}
\plotone{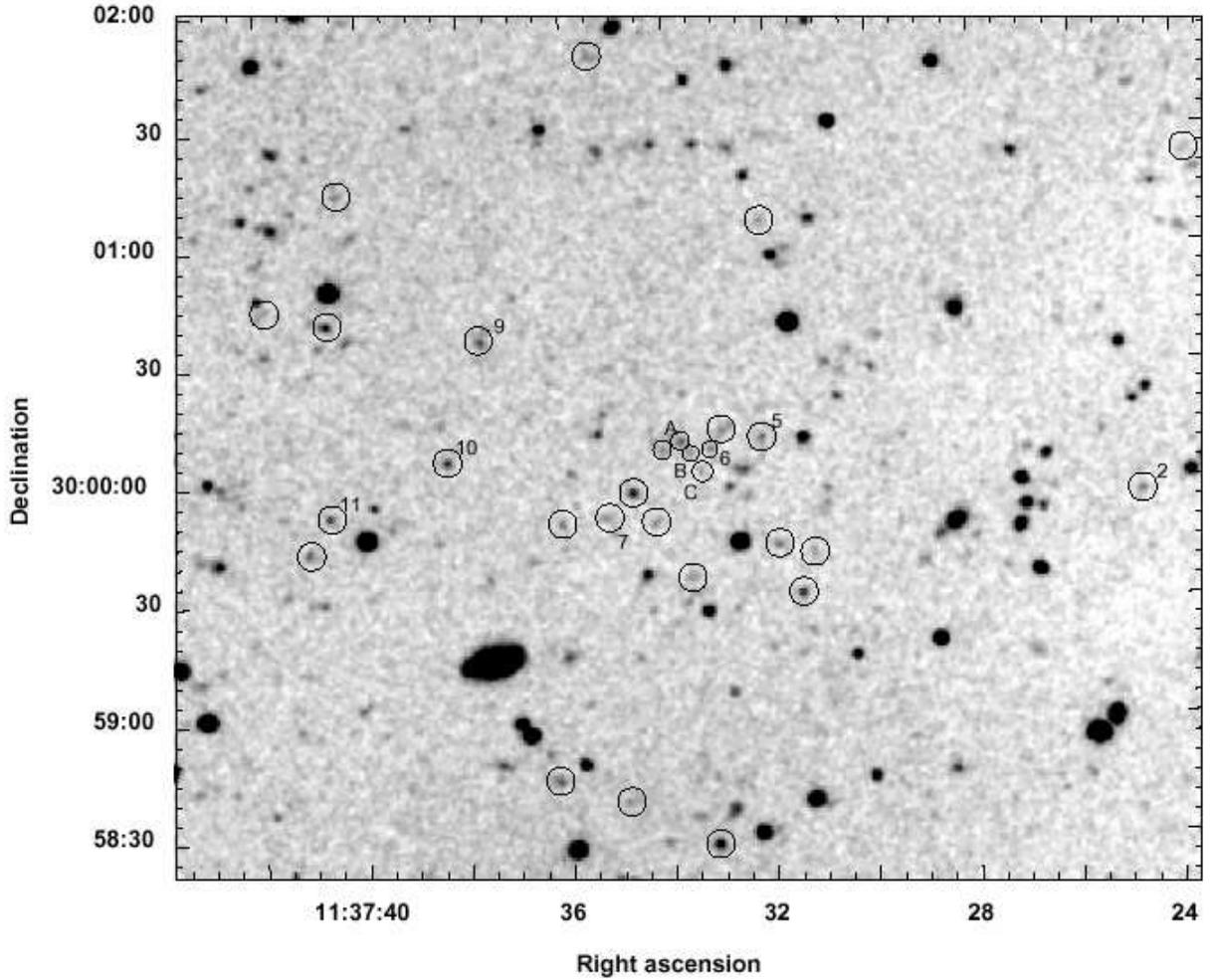}
\caption{$R$-band image from the MDM 1.3m telescope.  Objects with 
$4 < R-K < 6$ are circled, and are likely cluster members.  Spectroscopically
confirmed members are labeled with letters and numbers; details for these 
objects are given in Table 1.  The radio host galaxy is marked 'A'.  \label{fig:rimg}}
\end{figure}

\begin{figure}
\plotone{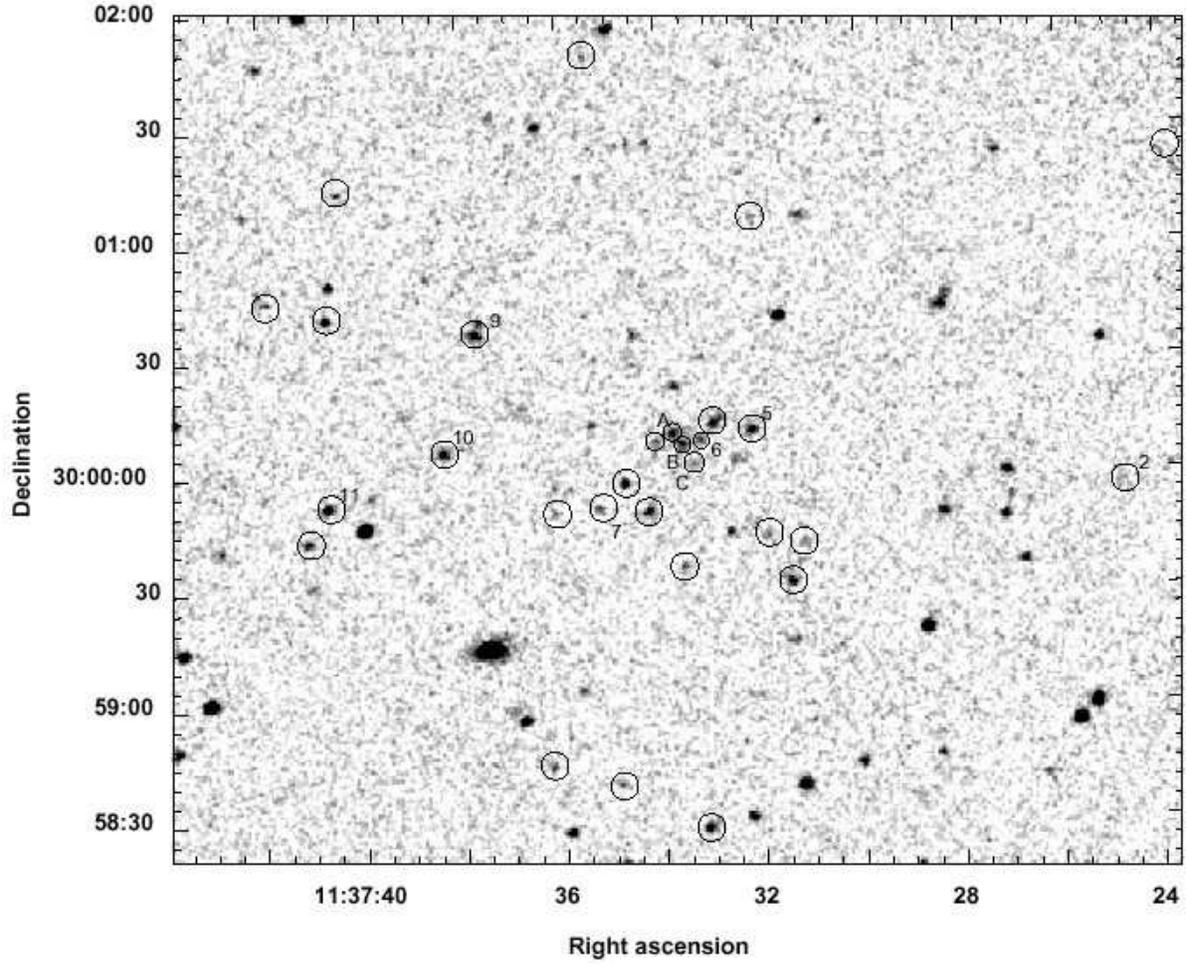}
\caption{$K$-band image from the KPNO 2.1m telescope with the IRIM detector. 
Image labels are as described in Figure~\ref{fig:rimg}. \label{fig:kimg}}
\end{figure}

The $R$-band and $K$-band images are presented in Figure~\ref{fig:rimg} and 
\ref{fig:kimg}.  Objects that were detected with at least 2.5$\sigma$ 
significance in both $R$ and 
$K$, with a minimum area of 12 pixels$^2$ (2.9 arcsec$^2$), and that have
$\rm{4 < (R-K) < 6}$ are circled.  Spectroscopically confirmed cluster
members (see \S\ref{sec:spect}) are labeled with letters and numbers; 
details for these objects are given in Table 1.  The radio source host 
galaxy is marked 'A'.  Source 'C' was detected at only $2\sigma$ in the 
$R$-band, and was not detected in the $J$-band, but is circled because it has
a measured redshift.

\section{Spectroscopy}\label{sec:spect}
\subsection{Long-slit Observation}
A long-slit observation (2 $\times$ 30 min.) of the cluster that included 
the radio counterpart and two
other galaxies was made at the Keck II 10m telescope with the Low Resolution
Imaging Spectrometer (LRIS; Oke et al.\ 1995) on 13 April
1998.  The spectra were corrected for bias and flat field effects 
using standard IRAF procedures.  They were wavelength calibrated using 
comparison Ne + Ar spectra and flux calibrated
using a standard star (Feige 34).  Redshifts were derived from 
cross-correlating the spectra with an elliptical galaxy template (M32
shifted to the rest frame).

\subsection{Multi-slit Mask Observation}
A multi-slit mask observation of the field surrounding the radio source
was made at the Keck II 10m telescope with LRIS on 29 May 2001.  The field
was observed for a total exposure of 60 min. (3 $\times$ 20 min.).  We reduced
the data from the LRIS-R, which was used with a 150 lines mm$^{-1}$ grating.
The central wavelength was $8495~\rm\AA$, the dispersion was 
$4.8~\rm\AA$/pix, and the pixel scale was 0.2109 arcsec/pix.
Objects were chosen for the slit mask based on their $(R-K)$ colors.  Objects
that had $4 < (R-K) < 6$ were candidates for inclusion on the mask, since
they provide us the best opportunity of confirming galaxies at redshifts
$z\approx1$.  The mask included a total of twelve program object slits.

The data were reduced following standard IRAF procedures.  The frames were
corrected for bias using the overscan region.  An internal halogen flat taken
through the mask following the data frames was used to correct for flat
field effects.  The three object frames were then aligned and combined with
cosmic rays rejected.

Spectra were successfully extracted for seven of the program objects; 
the remaining five were too faint for successful extractions.  
The spectra were wavelength-calibrated using comparison spectra of 
Hg + Ar lamps.  Redshifts were initially 
estimated by identifying a few obvious lines, including the Ca II H+K break,
and the [O II] $\lambda3727$ and [O III] $\lambda\lambda 4959,5007$
emission lines.  The spectra were then Fourier cross-correlated with an
elliptical galaxy template (M32 shifted to the rest frame) using FXCOR in
IRAF.  The errors in the redshifts are computed in FXCOR and are based on
the fitted peak height and the antisymmetric noise, or ``r-value'' (Tonry
\& Davis 1979).  The average (mean) error for the redshifts from both the 
long-slit and multi-slit mask observations is 150 km s$^{-1}$.

\subsection{Spectroscopic Results}
\begin{figure}
\plotone{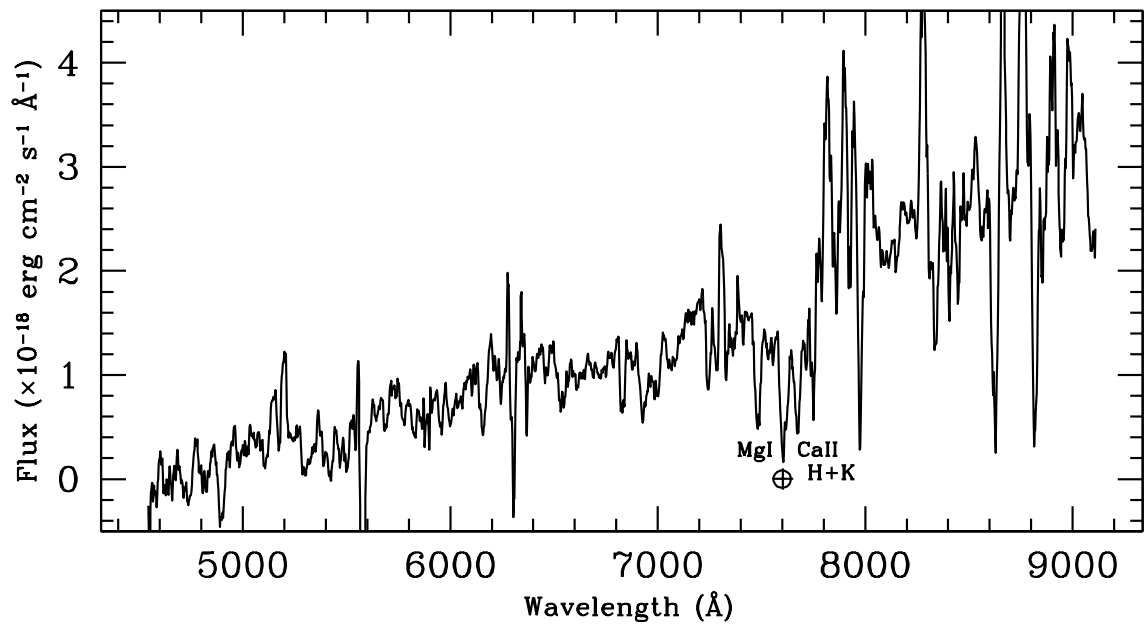}
\caption{Keck II LRIS spectrum of the radio source host galaxy.  The spectrum
has been smoothed with an 11 pixel boxcar.  The absorption features are
typical of an elliptical galaxy. \label{fig:spect}}
\end{figure}

All ten spectra that were successfully extracted had redshifts in the range
$0.950 < z < 0.965$, giving us a $100\%$ success rate of identifying cluster
galaxies based on their NIR/optical colors.  The objects and their
redshifts are listed in Table \ref{tab:phot}.  The spectrum of the radio host
galaxy is shown in Figure \ref{fig:spect}, and is typical of an elliptical
galaxy.  The spectrum has been smoothed with an 11 pixel boxcar.
We calculated the line-of-sight velocity dispersion using
$\sigma_{\|}=\sqrt{(N-1)^{-1}\sum_{i=1}^{N}\Delta v_i^2}$, where
$\Delta v_i=c(z_{i}-\bar z)/(1+\bar z)$.
The 68\% confidence uncertainty is calculated following Danese,
De Zotti, and di Tullio (1980), and includes errors due to the small number
of member galaxies in the field and the measurement errors.
We find $\sigma_{\|}=530^{+190}_{-90}$ km s$^{-1}$.  This is a typical
value for an Abell richness class 0 cluster (Lubin \& Bahcall 1993).

\section{Limit on the X-ray Luminosity}

An observation in the {\it ROSAT~} public data archive serendipitously
includes the position of BD1137+3000.  The {\it ROSAT~} PSPC observation was
centered at $\alpha$ = 11h36m33.6s, $\delta = +29^{\circ} 48\arcmin
00\farcs0$ leaving the position
of the bent-double radio source 17\farcm8 from the center, 
very close to the inner support structure of the PSPC.  The observation was 
performed on 1991 May 30, for a total duration of 33447 seconds.

We processed the data using the Snowden routines\footnote{``Cookbook for
Analysis Procedures for {\it ROSAT} XRT/PSPC Observations of Extended
Objects and the Diffuse Background'', S. L. Snowden (1994), with support from
the USRSDC} (Snowden et al.\ 1994)
which correct for exposure, vignetting, variations in the detector quantum
efficiency and non-cosmic background components.  After applying these
routines, 17884 seconds worth of good data remained.
We extracted the X-ray count rate within a $1\arcmin$ (480 kpc) radius 
aperture centered on the position of the radio source, and took background 
from two large circular apertures with similar offsets from
the center of the PSPC field as the source region.  The source region 
contained counts above the background at the $2.1\sigma$ level, with a $90\%$
confidence upper limit on the count rate of $1.1 \times 10^{-3}$ ct s$^{-1}$
in the $0.4 - 2.0$ keV energy band.
We converted this to a bolometric X-ray luminosity with PIMMS, using a 
model including Galactic absorption of $N_{H} = 1.93 \times 10^{20}$ erg
cm$^{-2}$ (Dickey \& Lockman 1990), a Raymond-Smith plasma with a 
temperature of 2.4 keV (estimated
from the $\sigma-T$ relation [Lubin \& Bahcall (1993)]), a chemical
abundance of 0.4 times the solar value, and a redshift of $z=0.96$.
The upper limit on the bolometric X-ray luminosity is then $L_{X,bol} \la
3 \times 10^{44}$ erg s$^{-1}$, which is consistent with the values expected
from the $L_{x} - \sigma$ relation ($3 \times 10^{44}$ erg s$^{-1}$ [Edge 
\& Stewart 1991]; $5 \times 10^{43}$ erg s$^{-1}$ [Borgani et al.\ 1999]).
{\it Chandra} and {\it XMM-Newton} observations are planned to confirm
the presence of an X-ray emitting ICM in this cluster, and to study it in
more detail.

\section{Conclusions}

We have provided evidence at optical and near-infrared wavelengths for a
high-redshift cluster of galaxies associated with a wide-angle tailed radio 
galaxy at $z = 0.96$.  Color-magnitude diagrams reveal an 
over-density of red objects near the radio source with colors consistent 
with elliptical galaxies at this redshift.  Objects are clustered on the
color-magnitude diagrams in the range $4 < (R-K) < 6$ and
$1.5 < (J-K) < 2.5$.

Using spectroscopy from the Keck II LRIS, we have
confirmed a total of ten galaxies
at the cluster redshift.  The line-of-sight velocity dispersion is
$\sigma_{\|}=530^{+190}_{-90}$ km s$^{-1}$, typical of an Abell
richness class 0 cluster.  Using data from the {\it ROSAT} public archive,
we limited the cluster's bolometric X-ray 
luminosity to $L_{X,bol} \la 3 \times 10^{44}$ erg s$^{-1}$, which is 
consistent with the value expected from the $L_{X}-\sigma$ relation.

The majority of high-redshift clusters found in traditional
X-ray and optical surveys select the most X-ray luminous and optically 
rich clusters.  The radio selection technique employed here is an important 
complement to those methods, in part because it can locate high-z clusters
with a wide range of X-ray luminosities and optical richnesses, thus aiding in
the study of galaxy evolution and cluster formation. 

E. L. B. thanks Adam Stanford and Craig Sarazin for helpful discussions 
regarding the NIR and X-ray data reduction, respectively.
Support for E. L. B. was provided by NASA through the {\it Chandra}
Fellowship Program, grant award number PF1-20017, under NASA contract number
NAS8-39073.  
Part of the work reported here was done at the Institute of Geophysics and
Planetary Physics, under the auspices of the US Department of Energy by
Lawrence Livermore National Laboratory under contract W-7405-Eng-48.
Support for M. D. G. was provided by a grant from the National Science 
Foundation (AST-99-70884).
The {\it FIRST} project is supported by grants from the National Geographic
Society, the National Science Foundation (AST 00-98259 and AST 00-98355), 
NASA (NAG 5-6035), NATO, IGPP, Columbia University, and Sun Microsystems.

\clearpage

\begin{deluxetable}{ccccccccc}
\tablewidth{0pt}
\tablecaption{Photometry and Redshifts for Confirmed Cluster Members \label{tab:phot}}
\tablehead{\colhead{id} & \colhead{RA(J2000)} & \colhead{Dec(J2000)} 
& \colhead{$m_R$} & \colhead{$m_J$} & \colhead{$m_K$} & 
\colhead{$J-K$} & \colhead{$R-K$} &
\colhead{z}}
\tablecomments{The redshifts for galaxies A, B, and C were determined from
the longslit observation; the redshifts for the remaining galaxies were
derived from the slitmask observation.  The radio source host galaxy is 
source A.  Source C was not detected in the $R$-band at 2.5$\sigma$, $m_R$ for
this galaxy is given at the 2$\sigma$ level.  It was not detected in the 
$J$-band.}
\startdata
A & 11:37:33.74 & 30:00:10.5 &  22.90 & 19.46 &  17.78 & 1.68 &5.12 & 0.9529\\
B & 11:37:33.59 & 30:00:07.6 &  23.57 & 20.17 &  17.90 & 2.27 &5.67 & 0.9524\\
C & 11:37:33.34 & 30:00:02.8 &  24.28 &\nodata& 18.91&\nodata&5.37 & 0.9551 \\
\hline
2 & 11:37:24.70 & 29:59:56.7 & 23.06 & 20.53  & 18.94  &1.59&  4.12 & 0.9591\\
5 & 11:37:32.16 & 30:00:11.4 & 22.92 & 19.83  &  17.71 &2.12&  5.21 & 0.9506\\
6 & 11:37:33.16 & 30:00:09.1 & 23.46 & 20.19  &  18.24 &1.95&  5.22 & 0.9571\\
7 & 11:37:35.18 & 29:59:51.5 & 23.38  & 20.65  & 18.78 &1.87&  4.59 & 0.9621\\
9 & 11:37:37.65 & 30:00:36.7 & 22.55  & 19.55  & 17.53 &2.02& 5.01 & 0.9582\\
10 & 11:37:38.34 & 30:00:06.2 & 22.43 & 19.23 & 17.67& 1.56 & 4.76 & 0.9556\\ 
11 & 11:37:40.63 & 29:59:52.5 & 22.83 & 19.40 & 17.42 &1.98&  5.42 & 0.9564\\
\enddata

\end{deluxetable}

\end{document}